\documentclass{iopart}
\usepackage{graphicx}
\newcommand{\op}[1]{{\hat{\mathcal#1}}}
\begin{document}
\title{Macroscopic optical response and photonic bands} 
\author{J.S. P\'erez-Huerta$^{1}$, Guillermo P. Ortiz$^2$, Bernardo
  S. Mendoza$^3$, and W. Luis Moch\'an$^1$} 
\address{
$^1$Instituto de Ciencias F\'{\i}sicas,
  Universidad Nacional Aut\'onoma de
  M\'exico,  Apartado Postal 48-3, 62251 Cuernavaca, Morelos, M\'exico\\
$^2$Departamento de
  F\'isica, Facultad de Cs. Exactas, Naturales y Agrimensura, Universidad Nacional del Nordeste, Av. Libertad 5400 Campus-UNNE, W3404AAS Corrientes, Argentina.\\
$^3$Division of Photonics,
  Centro de Investigaciones en Optica, \\Le\'on, Guanajuato,
  M\'exico\\
  E-mail: jsph: {\tt jsperez@fis.unam.mx}, wlmb: {\tt
    mochan@fis.unam.mx}, gpo: {\tt gortiz@exa.unne.edu.ar},
    bsm: {\tt bms@cio.mx}
}

\begin{abstract}
  We develop a formalism for the calculation of the macroscopic
  dielectric response of composite systems made of particles of one
  material embedded periodically within a matrix of another material,
  each of which is characterized by a well defined dielectric
  function. The nature of these dielectric functions is arbitrary, and
  could correspond to dielectric or conducting, transparent or opaque,
  absorptive and dispersive materials. The geometry of the particles
  and the Bravais lattice of the composite are also arbitrary. Our
  formalism goes beyond the longwavelenght approximation as it fully
  incorporates 
  retardation effects. We test our formalism through the study the
  propagation of 
  electromagnetic waves in 2D photonic crystals 
  made of periodic arrays of cylindrical holes in a dispersionless
  dielectric host. Our macroscopic theory yields
  a spatially dispersive macroscopic response which allows the
  calculation of the full photonic band structure of the system, as
  well as the characterization of its normal modes,  upon
  substitution into the macroscopic field equations. We account
  approximately for the spatial dispersion through a local magnetic
  permeability and we analyze the resulting dispersion relation,
  obtaining a region of left-handedness. 

\end{abstract}

\section{Introduction}

The propagation of light within homogeneous materials
is completely characterized by their electromagnetic
response. In the optical regime, magnetic effects
are typically negligible so that knowledge of the dielectric 
 response is sufficient for the study of wave 
propagation \cite{Bredov}. However,
when the system has spatial inhomogeneities,
the behaviour of light is not trivial and has 
captured the attention of many researchers 
\cite{Notomi, Barrera, Yeh,Mochan,Mochan2,Kosaka,Stroud,Ortiz(2001),Ortiz(2003)a,Hilke}. 
The possibility of controlling the propagation of photons by producing
artificially structured  materials has been widely demonstrated in
recent years. Novel effects have been obtained, such as negative
refraction, inverse Doppler effect, optical invisibility cloaking, 
optical magnetism  
etc. \cite{Chen,Zhang,Lezec,Urzhumov}. 

Effective medium theories
\cite{Milton,Bergman,Tao,Datta,Alexopoulos,Doyle,Ludwig,Costa,Ortiz(2003)b} 
have been proposed to describe inhomogeneous
systems, such as diluted colloidal suspensions
in the quasistatic or long wavelength limit, in terms of a
homogeneous macroscopic response.
Further developments in homogenization of composites have been proposed
\cite{Silveirinha,Mochan3,Myroshnychenko,Guenneau,Cabuz,Ortiz,Raghunathan,haleviprl,halevip}
 and their limits of validity have have
been discussed 
\cite{Notomi, Simovski,Menzel}. 

In this paper we obtain  the macroscopic 
optical response using a computationally efficient reformulation of
a procedure \cite{Mochan,Mochan2} original developed to account for
the local field effect 
of  systems with spatial fluctuations.  
Our main result is that for any system, the macroscopic response   
to an {\em external} excitation is simply the average projection of the
corresponding microscopic response. We apply this
general result to Maxwell equations in order to obtain an explicit expression
for the  macroscopic dielectric response of binary composites. We illustrate
its use with numerical results for a 2D periodic photonic crystal.
As the wavelength of the electromagnetic field becomes comparable to
the characteristic microscopic spatial lengthscale of the
system, the macroscopic response becomes highly non-local, with a
strong dependence on the wavevector besides its usual dependence on
frequency. When the full spatial dispersion is taken into account, the
photonic band structure of the system can be obtained and analyzed from its
macroscopic response. 

The paper is organized as follows: 
In Sec. \ref{Theory} we formulate a  general 
theory for the macroscopic dielectric response 
of composite systems. In Sec. \ref{binary} we adapt our
formulation to periodic systems made of two alternating materials,
each characterized by 
well defined dielectric functions. In Sec. \ref{HaydockMethod} we
develop an efficient computational scheme
which allows the numerical calculation of the macroscopic response
without recourse to explicit operations between large matrices.
In Sec. \ref{Numeric} we obtain the
non-local macroscopic dielectric tensor and use it to calculate
macroscopically the
full band structure of  a 2D photonic crystal
made up of non-dispersive dielectric components;  we compare it to exact
results as well as to results within a local approximation that
partially accounts for spatial dispersion through an effective
magnetic permeability. Finally, in
Sec.~\ref{conclusion} we present our conclusions. 

\section{General theory}
\label{Theory}
Consider a non-magnetic inhomogeneous system characterized by its
dielectric response $\hat\epsilon$, defined through the constitutive
equation $\mathbf D =\hat\epsilon\mathbf E $, where $\mathbf D$ and $\mathbf
E$ are the {\em microscopic} displacement and electric fields
respectively. We designate these fields as microscopic as they have
spatial fluctuations due to the inhomogeneities of the material. They
are not microscopic in the atomic scale, but rather, on the scale of
the spatial inhomogeneities of the system. We use a caret ($\,\hat{}\,$)
over a symbol to denote its operator nature and we leave implicit the
dependence of the dielectric response on position as well as on
frequency.
Our
purpose is to obtain the macroscopic dielectric response of the system, relating
the macroscopic displacement and electric fields, from which the
fluctuations have been removed.  

We start from Maxwell equations for monochromatic microscopic
fields. We 
follow the usual procedure \cite{Jackson} to decouple the magnetic field from the
electric field to obtain 
the second order {\em wave} equation
\begin{equation}\label{weq} 
\op W \mathbf E =\frac{4\pi c }{i q}\mathbf J
\end{equation}
where 
\begin{equation}\label{Waveop}
  {\op W}=\hat\epsilon-\frac{1}{q^2}\nabla\times\nabla\times =
  \hat\epsilon+\frac{1}{q^2}\nabla^2\op P^T  
\end{equation}  
is the {\em wave operator}, $q=\omega/c$ is the wavenumber of light in vacuum, 
$\omega$ is the frequency, $c$ is the speed of
light in vacuum and $\mathbf J$ is the {\em external} electric current
density. Here we introduced the transverse projector $\op P^T$ such that
the transverse projection of a field $\mathbf F$ is $\mathbf
F^T=\op P^T\mathbf F$. For completeness, we also introduce the
longitudinal projector $\op P^L$ such that $\op P^L+\op P^T=\hat 1$, with
$\hat 1$ the identity operator. Notice that Eq. (\ref{weq}) contains both a
longitudinal and a transverse part, so it describes not only transverse
waves, but also allows for the possible excitation of longitudinal waves such
as plasmons \cite{Mochan4}.

We solve Eq.~(\ref{weq}) 
formally for $\mathbf E $  to obtain
\begin{equation}\label{solvE}
  \mathbf E =\frac{4\pi c }{i q}{\op W}^{-1}\mathbf J
  .
\end{equation}
As we are interested on the macroscopic response of the system, we
introduce the average 
$\op P_a$ and fluctuation $\op P_f$ projectors, such that an arbitrary field
$\mathbf F$ can be written as $\mathbf F=\mathbf F_a+\mathbf F_f$ with $\mathbf
F_a=\op P_a \mathbf F$ its average and $\mathbf F_f=\op P_f
\mathbf F$ its fluctuating part, and we identify the average field
with the
macroscopic field $\mathbf F_a\equiv\mathbf F_M$. Later, we will
provide appropriate explicit definitions for 
$\op P_a$ and $\op P_f$; here we remark that they must be
idempotent $\op P_a^2=\op P_a$, $\op P_f^2=
\op P_f$, i.e., the average of the average is the average, and the
fluctuations of the fluctuations are the fluctuations \cite{Mochan}. 
Furthermore, 
$\op P_a\op P_f=\op P_f\op P_a=0$ and $\op P_a + \op P_f=\hat 1$. 

As the macroscopic response would be
useless unless the {\em external excitations are devoid of
microscopic spatial fluctuations}, we assume
that the external current has no fluctuations, so $\mathbf J_f=\mathbf
0 $ and $\mathbf J= \mathbf
J^M=\op P_a \mathbf J^M$, where we used the idempotency of $\op P_a$. 
Thus, acting on both sides of Eq.~(\ref{solvE}) with $\op P_a$ we obtain
\begin{equation}  \label{Eaverage} 
  \mathbf E^M=\frac{4\pi c}{i q} \op W^{-1}_{aa}\mathbf
  J^M.
\end{equation}
Here we define $\op
O_{\alpha\beta}\equiv \op P_\alpha\op O\op P_\beta$, with
$\alpha,\beta=a,f$ for any operator $\op O$. As Eq. (\ref{Eaverage}) 
relates the macroscopic external electric current to the macroscopic
electric field we may identify the macroscopic inverse wave operator
$\op W_M^{-1}$,
\begin{equation}\label{Emacros}
  \mathbf E^M=\frac{4\pi c }{i q} \op W^{-1}_M\mathbf J^M
\end{equation}
where
\begin{equation}\label{WaveMaMi}
  \op W^{-1}_M=\op W^{-1}_{aa}.
\end{equation}
We summarize this results stating that the macroscopic inverse wave
operator is simply the average of the microscopic inverse wave
operator. This is a particular case of a more general result: the
macroscopic response to an external excitation is simply the average
of the corresponding microscopic response.

From the macroscopic Maxwell equations we may further relate
the macroscopic wave operator in Eqs. (\ref{Emacros}) and (\ref{WaveMaMi})  
to the macroscopic dielectric response
of the system $\hat\epsilon^M$ through
\begin{equation}\label{MWaveop}
  \op W^M=\hat\epsilon^M+\frac{1}{q^2}\nabla^2\op P_a\op P^T,
\end{equation} 
in analogy to Eq. (\ref{Waveop}).
Thus, we
have to invert the wave operator, average it, and invert it again to
finally identify the
macroscopic dielectric response. Notice that we have made no
approximation whatsoever. We remark that as $\hat\epsilon$ relates two fields,
$\mathbf E$ and $\mathbf D$, that
have spatial fluctuations, we may not simply average $\hat\epsilon$ to obtain
its macroscopic counterpart, i.e. $\hat\epsilon^M
\ne\hat\epsilon_{aa}$. The difference constitutes the {\em local
  field} effect \cite{Mochan}.

Our result may
easily be generalizable to other situations and other response
functions. The procedure consists on first identifying the response
(in our case $\op W^{-1}$) to
the {\em external} perturbation (i.e. $\mathbf J$), and then
averaging it to yield the 
corresponding macroscopic response (i.e. $\op W^{-1}_M=\op
W^{-1}_{aa}$), which may further be related to 
the {\em desired} macroscopic response operator (i.e.
$\hat\epsilon^M$). 

\section{Periodic binary systems}
\label{binary}

In this section we use Eq.~(\ref{WaveMaMi})
to obtain the optical properties of an artificial   
binary crystal made of two materials $A$ and $B$ with dielectric
functions  $\epsilon_A$ and $\epsilon_B$. We assume that both media are local
and isotropic so that $\epsilon_A$ and $\epsilon_B$ are simply complex
functions of the frequency. For convenience, we will further assume
that $\epsilon_A$ is real, though this assumption is easily relaxed
\cite{unpublished}.

We introduce the characteristic
function $B(\mathbf r)$  of the inclusions, such that $B(\mathbf r)\equiv1$
whenever $\mathbf r$ is on the region $B$ 
occupied by the inclusions, and  $B(\mathbf r)\equiv0$ otherwise. Thus, we
may write the microscopic dielectric response as 
\begin{equation}
  \label{epsmicro}
  \epsilon(\mathbf r)= 
  \frac{\epsilon_A}{u}\left( u-B(\mathbf r) \right),
\end{equation}
where we defined the spectral variable $u\equiv
1/(1-\epsilon_B/\epsilon_A)$ \cite{Bergman}. The microscopic wave operator 
of Eq.~(\ref{Waveop}) may be written as
\begin{equation}
  \label{Waveopbin}
  \op W= 
  \frac{\epsilon_A}{u}\left(  u\hat g^{-1} -
  \op B\right), 
\end{equation}
where the characteristic operator $\op B$ corresponds to
multiplication by $B(\mathbf r)$ in real 
space, and we defined
\begin{equation}
  \label{metric}
  \hat g=\left(1+\frac{\nabla^2\op P^T}{q^2\epsilon_A}\right)^{-1}.
\end{equation}
which, as shown below, plays the role of a metric. 

Using Eq.~(\ref{Waveopbin}) we write Eq.~(\ref{WaveMaMi}) as 
\begin{equation}
  \label{WaveMacrog}
   \op W^{-1}_{M}=
   \frac{u}{\epsilon_A}\hat g_{aa}\left( u -
   \op B\hat g\right)^{-1}_{aa},
\end{equation}
where we have taken advantage of the fact that $\hat g$ is unrelated to the
texture of the crystal, so that it does not couple average
to fluctuating fields. 

Using Bloch's theorem \cite{Sakoda}, we consider an electric field of
the form \cite{Datta} 
\begin{equation}
  \mathbf E _\mathbf k (\mathbf r)=\sum_{\mathbf G} \mathbf E _{\mathbf
    G}e^{i(\mathbf k +\mathbf G)\cdot\mathbf r },
  \label{ElectricFB}
\end{equation}
 where $\mathbf k $ is a given wavevector, $\{\mathbf G\}$ is  the 
 reciprocal lattice  of our crystal and the coefficients $\mathbf E _{\mathbf
    G}$ represent the field in  reciprocal space. In this representation,
 all operators may be written as matrices with vector index pairs
 $\mathbf G,\mathbf G'$, besides other possible indices, such as
 Cartesian ones.
Thus, we represent the 
 longitudinal projector as the matrix 
 \begin{equation}
   \mathcal P^L_{\mathbf  G\mathbf G'}= \delta_{\mathbf G\mathbf G'}
   \frac{(\mathbf k +\mathbf
     G)}{|\mathbf k +\mathbf G|}  
   \frac{(\mathbf k +\mathbf G')}{|\mathbf k +\mathbf G'|}
   \label{LongP}
 \end{equation}
  with $\delta_{\mathbf G\mathbf G'}$ the Kronecker's delta, so the
  transverse projector becomes $\mathcal P^T_{\mathbf  G\mathbf
    G'}=\mathbf 1 \delta_{\mathbf G\mathbf G'}-\mathcal P^L_{\mathbf
    G\mathbf G'}$ with $\mathbf 1$ the Cartesian identity matrix.
 The Laplacian operator in reciprocal space is represented by 
 \begin{equation}
   \nabla^2\to -(\mathbf k +\mathbf G)^2 \delta_{\mathbf G\mathbf G'},
   \label{Laplace}
 \end{equation}
and we can define the average projector as a cutoff in the reciprocal
space \cite{Mochan2}
\begin{equation}
  (\mathcal P_a)_{\mathbf G\mathbf G'}=\delta_{\mathbf G\mathbf
      0}\delta_{\mathbf G'\mathbf 0 },  
  \label{AverageP}
\end{equation}
so that average fields simply keep the contributions with vector
$\mathbf k$ while 
all other wavevectors are filtered out.
 
As shown by Eq.~(\ref{WaveMacrog}),  we only require
\begin{equation}\label{G00}
  (u -\op B\hat g)^{-1}_{aa}=(u -\op B\hat g)^{-1}_{\mathbf 0\mathbf 0}.
\end{equation}
to obtain the macroscopic inverse wave operator, where the subindices
$\mathbf 0$ denote the projection onto the subspace with $\mathbf
G=\mathbf 0$.

\section{Recursive method }
\label{HaydockMethod}

The calculation of Eq.~(\ref{G00}) is analogous
to that of a projected Green's function  \cite{Sutton,DattaB} 
\begin{equation}
  {\cal G}_{aa}(\varepsilon)  =
  \langle a |\op G (\varepsilon)| a\rangle,
  \label{GreenProj}
\end{equation}
onto a given state $|a\rangle$, where
\begin{equation}
  \op G(\varepsilon)=(\varepsilon -\op H)^{-1}
  \label{Green}
\end{equation}
is the Green's operator corresponding to some Hermitian Hamiltonian
$\op H$ and $\varepsilon$ is a complex energy.  
In Ref. \cite{Mochan3} a similar result was obtained, where the
Hamiltonian $\op H$ was identified with the longitudinal projection of the
characteristic function $\op B^{LL}$, the energy $\varepsilon$ with the
spectral variable $u$, and $|a\rangle$ with a slowly varying
longitudinal wave, and it was shown that the corresponding projected
Green's function 
was proportional to the inverse of the longitudinal macroscopic
dielectric function. 
Haydock's method may be applied to obtain the projected Green's function
(\ref{GreenProj}) in a  
very efficient way \cite{Haydock,Haydock2}, and it  has been adapted previously 
\cite{Mochan3} to the calculation of the optical response of nanostructured
systems in the long-wavelength local limit. 
In this  section we
generalize the approach of Ref. \cite{Mochan3} to arbitrary
frequencies  and wavevectors.

Eqs.~(\ref{GreenProj}) and (\ref{Green})  are similar to 
Eq.~(\ref{G00}), although  the operator product $\op H=\op B\hat g$ that plays
the role of Hamiltonian
does not correspond to a symmetric matrix. Nevertheless, it would
correspond to a 
Hermitean operator if we use $\hat g$ as a metric, that is, if we
use $\langle\psi|\hat g|\phi\rangle$ instead of
$\langle\psi|\phi\rangle$ as the scalar product of two
arbitrary states $|\psi\rangle$ and $|\phi\rangle$. Then, it is easy
to verify the Hermiticity condition $[\langle\psi|\hat g(\op
H|\phi\rangle)]^*=\langle\phi|\hat g(\op H|\psi\rangle)$. Notice,
however, that $\hat g$ is not positive definite. 

According to Eq.~(\ref{WaveMacrog}), we need the average
$(u -\op B\hat g)^{-1}_{aa}\to\langle a|(u -\op B\hat g)^{-1}|a\rangle$, where
we projected onto an {\em average} state $|a\rangle=b_0|0\rangle$
consisting of a plane wave with a given wavevector $\mathbf k$,
frequency $\omega$ and polarization $\mathbf e$. Here, $|0\rangle$ is
normalized according to the metric $\hat g$, i.e., $\langle
0|\hat g|0\rangle=g_0=\pm 1$, and the coefficient $b_0$ is a real
positive number chosen to 
guarantee that $|a\rangle$ is normalized in the usual sense $\langle
a|a\rangle=1$. 
Now, we define $|-1\rangle\equiv0$ and we  obtain new states by means of
the recursion relation 
\begin{equation}\label{n+1}
  |\widetilde{n+1}\rangle\equiv \op B\hat g|n\rangle = b_{n+1}|n+1\rangle + a_n
  |n\rangle +  g_{n-1} g_n b_{n}|n-1\rangle, 
\end{equation}
where all the states $|n\rangle$ are orthonormalized
according to the metric
$\hat g$, that is 
\begin{equation}\label{ortonor}
  \langle n|\hat g|m\rangle = g_n \delta_{nm},
\end{equation}
with $g_n=\pm1$ and $\delta_{nm}$ the Kronecker's delta
function. The requirement of orthonormality yields the generalized
Haydock coefficients $a_n$, $b_{n+1}$, and $g_{n+1}$ given the
previous coefficients $b_{n}$, 
$g_{n}$ and $g_{n-1}$. Thus,  $a_n$ are obtained from
\begin{equation}\label{an}
  \langle n|\hat g|\widetilde{n+1}\rangle = a_n g_n,
\end{equation}
and $b_{n+1}$ from
\begin{equation}\label{bn}
  \langle \widetilde{n+1}|\hat g|\widetilde{n+1}\rangle = g_{n+1} b_{n+1}^2 
  + g_n a_n^2 + g_{n-1} b_{n}^2, 
\end{equation}
where we choose the sign 
$g_{n+1}=\pm1$ so that $b_{n+1}^2$ is positive and we may choose
$b_{n+1}$ as a real positive number. In the
basis $\{|n\rangle\}$, $\op B\hat g$ is {\em represented} by a tridiagonal matrix
with $a_n$ along the main diagonal, $b_n$ along the subdiagonal and
$ g_{n-1} g_n b_n$ along the supradiagonal, so that
\begin{equation}
\label{MWaveMatrix}
u-\op B\hat g \to \left(
\begin{array}{ccccc}
 u-a_0 & -b_1g_1g_0 & 0  & 0&\cdots\\
 -b_1 & u-a_1 & -b_2g_2g_1 & 0& \cdots\\
 0   & -b_2 & u-a_2 & -b_3g_3g_2& \cdots\\
\vdots&\vdots&     &\vdots &\ddots
\end{array}
\right). 
\end{equation}
Notice that we may represent the states $|n\rangle$ through the
corresponding 
Cartesian field components for each reciprocal vector $\mathbf G$ or
for each position $\mathbf r$ within a unit cell. Thus, the action of
$\hat g$ is a trivial multiplication in reciprocal space, while that
of $\hat\mathcal B$ is a trivial multiplication in real space, and we
may repeatedly compute $\hat \mathcal B\hat g|n\rangle$ and Haydock's
coefficients {\em without performing any actual
matrix multiplication}, by alternatingly fast-Fourier transforming our
representation between real and reciprocal space.

According to Eq. (\ref{WaveMacrog}), we do not require the full
inverse of the matrix (\ref{MWaveMatrix}), but only the element in the
first row and first column. Following Ref. \cite{Mochan3}, we obtain
that element as a continued fraction, which substituted into
Eq. (\ref{WaveMacrog}) yields
\begin{equation}
\label{WMacroFC}
  \op W^{-1}_M \to \sum_{ij}e_i (\mathcal W^{-1}_M)_{ij} e_j=
  \frac{u}{\epsilon_A}
  \frac{g_0 b_0^2}
       {u-a_0 
         -\frac{g_0 g_1 b_1^2}
         {u -a_1 
           -\frac{g_1 g_2 b_2^2}
               {u- a_2 
                 -\frac{g_2 g_3 b_3^2}{\ddots
       }}}}.
\end{equation}
The right hand side of Eq. (\ref{WMacroFC}) denotes the macroscopic
response projected onto a state with the given wavevector $\mathbf k$,
frequency $\omega$ and polarization $\mathbf e$. Due to the
translational invariance of the homogenized system, the inverse wave operator
for a given $\mathbf k$ corresponds simply to a tensor with  Cartesian
components $(\mathcal W^{-1}_M)_{ij}$. Thus, so far we have calculated
its inner products with $\mathbf e$, as shown by the second term of
Eq. (\ref{WMacroFC}). We
may repeat the calculation above for different orientations of the
(possibly complex) unit polarization vector $\mathbf e$ and solve the
resulting equations 
for {\em all} the individual Cartesian components $(\mathcal W^{-1}_M)_{ij}$.
Finally, we perform a matrix inversion of this macroscopic tensor and
we use Eq.  
\ref{MWaveop} and \ref{WaveMaMi} to compute the 
macroscopic dielectric tensor
\begin{equation}
  \label{EpsMacroTensor}
  \epsilon^M_{ij}(\omega,\mathbf k)=
  \frac{1}{q^2}(k^2\delta_{ij}- k_i  k_j)+\mathcal W^M_{ij}(\omega,
  \mathbf k).
\end{equation}
Eqs. (\ref{WMacroFC}) and (\ref{EpsMacroTensor}) constitute the main
result of our formalism. Notice that in general, this dielectric
tensor would depend on both $\omega$ and $\mathbf k$, that is, it is
a temporal and spatially dispersive response function.

\section{Results}
\label{Numeric}

To test our formalism, we first calculate the
local limit of the longitudinal macroscopic
dielectric function
$\epsilon^M_L(\omega)\equiv\epsilon^M_{xx}(\omega,\mathbf k\to0 \mathbf
x)$ of a 2D 
system made up of a square array of empty
cylindrical holes ($\epsilon_B=1$) placed with lattice parameter $a$
within a dielectric medium with permittivity $\epsilon_A=12$. The
holes are of  circular cross section with
radius $\rho=0.45a$. We assume that the axes of
the cylinders are parallel to the $z$ axis. This system has been
frequently been used as a testground for calculations of photonic
crystal properties \cite{Busch}.
\begin{figure}
\includegraphics{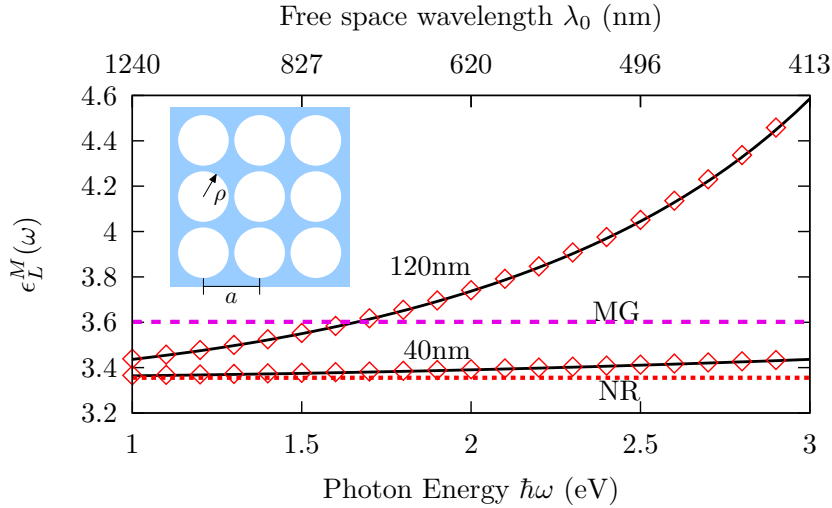}
\caption{(Color online) Longitudinal macroscopic dielectric function
  $\epsilon^M_L(\omega)$ of a 2D 
system (inset) made up of a square array with lattice parameter $a$ of empty
cylindrical holes ($\epsilon_B=1$) of circular cross section with
radius $\rho=0.45a$  within a dielectric
medium with permittivity $\epsilon_A=12$. We show the results of using
Maxwell-Garnett formula (MG), of a nonretarded calculation (NR), of a
numerically exact matrix calculation (diamonds) and of our theory
(solid lines) for two different lattice parameters $a=40$nm and $120$nm.
 }
\label{fig:Eps_M_L1L2}
\label{f1}
\end{figure}
In this calculation we took the limit $\mathbf k\to\mathbf0$ along the
$x$ direction, and we introduced the Cartesian unit vectors $\mathbf x$, 
$\mathbf y$ and $\mathbf z$. We remark that it is important
to assign a direction to $\mathbf k$ even in this limit, as the
transverse $T$ and longitudinal $L$ projection of $\mathcal W^M$ 
differ, although, for this system, the resulting
$\epsilon^M(\omega)=\epsilon^M(\omega,\mathbf k\to \mathbf 0)$ is
isotropic within the $xy$ plane.  

\begin{figure}
  \includegraphics{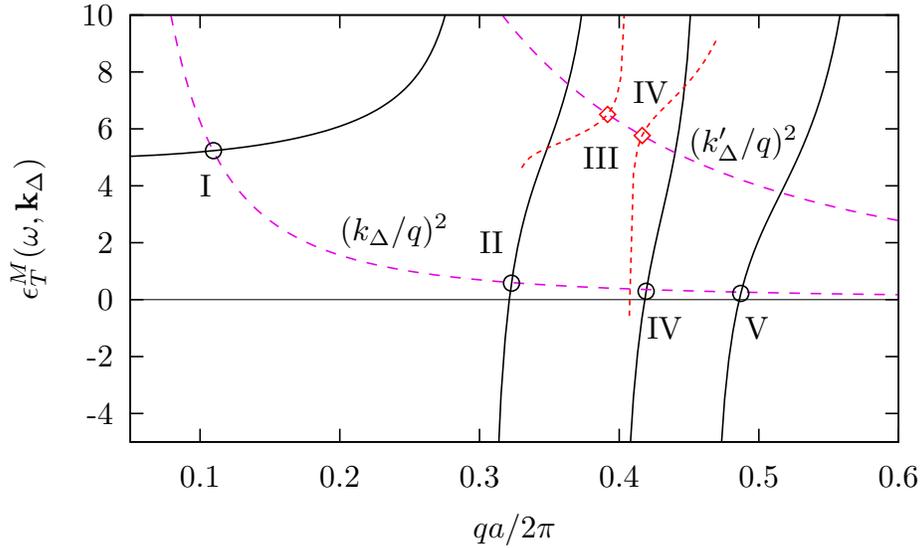}
  \caption{(Color online) Out-of-plane macroscopic dielectric function
    $\epsilon^M_{T}(\omega,\mathbf k_\Delta)$ of the same
    system as in Fig. \ref{fig:Eps_M_L1L2} for a given wavevector
    $\mathbf k_\Delta$ along the $\Delta$ line of the first
    Brillouin zone (solid lines). The line
    $(k_\Delta/q)^2$ (long dashes) intersects (circles)
    $\epsilon^M_{T}(\omega,\mathbf k_\Delta)$ at the
    frequencies $\omega_\alpha(\mathbf k_\Delta)$ of the TE normal
    modes, with $\alpha=$I, II, IV, V. The intersections (diamonds)
    of $\epsilon^M_{T}(\omega,\mathbf k'_\Delta)$ (short dashes) with
    $(k'_\Delta/q)^2$ (long dashes) corresponding to a shifted
    wavevector $\mathbf k'_\Delta=\mathbf 
    k_\Delta+(2\pi/a)\mathbf y$ yields the modes $\alpha=$III, IV (see text).}
  \label{fig:TE}
  \label{f2}
\end{figure}

In Fig. \ref{fig:Eps_M_L1L2} we 
show  $\epsilon^M_L(\omega)$  calculated for two different 
lattice parameters $a=40$nm and $a=120$nm. Our calculations were
performed using the Perl Data Language \cite{PDL} \footnote{Available from http://pdl.perl.org} .
 The figure shows that our
formalism yields the same results as a full straightforward matricial
calculation \cite{Ortiz} that solves Maxwell equations in a plane wave
basis. Nevertheless, our calculation is much more economical in memory
usage, as we don't have to store the full matrices that represent the
dynamics of the system, and is at least four orders of magnitude
faster as we do not perform any matrix manipulation with our
scheme. To illustrate the effects of retardation, in the figure we
also show the result of using  the non-retarded calculation
of Ref. \cite{Mochan3} (NR), together with those of the
well-know 2D Maxwell-Garnet (MG)
effective medium formula \cite{Datta}. As the dielectric functions of the
components are independent of the frequency, since the constituents were
taken to be non-dispersive, in this case the frequency dependence of the result
arises solely from to the finite ratio of the free space wavelength to
the lengthscale of the system, namely, the lattice parameter
$a$. Thus, in the  $\omega\to 0$ limit our results
approach those of the NR calculation. The difference
between the NR and MG calculations is due to the high filling fraction
$f\approx0.64$ of the cylinders, which interact non-dipolarly with
their nearest neighbors which they almost touch, while MG contains
only dipolar interactions. Naturally, retardation effects become
stronger as $a$ increases. Furthermore,
$\epsilon^M(\omega,\mathbf  
k)$ depends only on the products $qa$ and $\mathbf k a$ so that the
two curves shown in Fig. \ref{f1} could be collapsed into one curve if
appropriately plotted. We use this in Fig. \ref{f2}, which
shows the transverse response $\epsilon^M_T(\omega, \mathbf
k_\Delta)=\epsilon^M_{zz}(\omega, \mathbf k_\Delta)$  
corresponding to out-of-plane or TE polarization as a  function of the
normalized frequency $qa/2\pi$. Here, we have chosen  a 
finite in-plane  wavevector  $\mathbf k_\Delta =(\pi/2a,0,0)$,
corresponding to the midpoint 
of the $\Delta$ line between $\Gamma$ and $X$ in the first Brillouin
zone (BZ) \cite{Sakoda}. This figure covers
a larger frequency range than Fig. \ref{f1}, and therefore it displays
a series of poles related to resonant multiple coherent reflections at
the interfaces between the A and B regions. 

The dispersion relation of transverse waves with TE polarization
propagating along the $xy$ plane can be simply written as
\cite{Landau}
\begin{equation}\label{TEP}
  \epsilon^M_{T}(\omega,\mathbf k)=\frac{k^2}{q^2}.
\end{equation}
Therefore, in Fig. \ref{f2} we also plotted the curve $(k_\Delta/q)^2$. Those
points where it intersects the curves for $\epsilon^M_{T}$ correspond
to normal TE modes of the system with frequencies
$\omega_\alpha(\mathbf k=\mathbf k_\Delta)$,  $\alpha=$I, II, \ldots
Here, we use roman numerals to number the modes in increasing frequency order. 

\begin{figure}
  \includegraphics{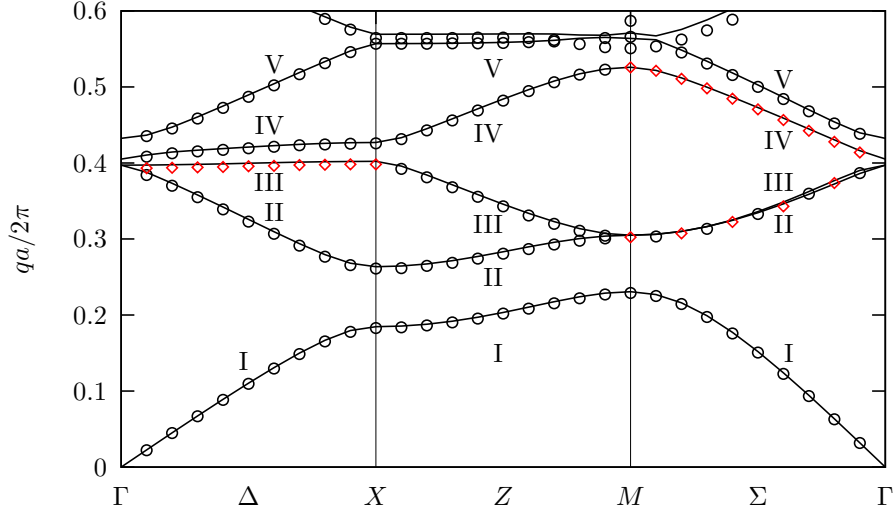}
  \caption{(Color online) Photonic bands of the TE modes of the same
    system as in 
    Fig. \ref{f1} obtained by solving Eq. (\ref{TEP}) (circles). 
For
    comparison, we also show the photonic bands obtained from the
    solution of 
the corresponding eigenvalue problem, 
    Eq. (\ref{TEEP}),
 as described in
    the text (solid lines).
 We also show the modes obtained by
    shifting the wavevector $\mathbf k\to\mathbf k'=\mathbf k +
    (2\pi/a)\mathbf y$ (diamonds).}
  \label{f3}
\end{figure}

In Fig. \ref{f3} we show the photonic bands (circles) of the TE modes of our
system obtained by solving the dispersion relation (\ref{TEP}) as
$\mathbf k$ travels along the line $\Gamma-\Delta-X-Z-M-\Sigma-\Gamma$
within the first BZ.  
In order to test our theory, we also plot the TE modes obtained by
solving the eigenvalue equation \cite{Joannopoulos}
\begin{equation}\label{TEEP}
  \left(\frac{1}{\sqrt{\epsilon(\mathbf
      r)}}\nabla^2\frac{1}{\sqrt{\epsilon(\mathbf
        r)}}\right) \left(\sqrt{\epsilon(\mathbf r)}E_z(\mathbf r)\right) =
  -\frac{\omega^2}{c^2}\left(\sqrt{\epsilon(\mathbf r)}E_z(\mathbf r)\right)
\end{equation} 
through an implicitly restarted Arnoldi
iteration \cite{Saad} , applying $1/\sqrt{\epsilon(\mathbf r)}$
in real space and
$\nabla\to i(\mathbf k+\mathbf G)$ in reciprocal space \cite{unpublished}. We further
verified our results by recalculating them with the freely available package MPB
 \cite{MIT} and comparing it to previous results
\cite{Busch} for the same system. The agreement between
these calculations shows that it is feasible to use the macroscopic
dielectric response of the system to obtain its photonic band
structure. However, to succeed, we have to account fully for the
spatial dispersion of $\epsilon^M_{T}(\omega,\mathbf k)$. 

We remark that although we obtained the correct photonic bands even for
large wavevectors, going all the way around the first BZ, there
are some regions where we failed to obtain all of the normal modes. For
example, our procedure did not produce the third band (labeled III in
the figure) along
the $\Delta$ line, and neither the fourth band (IV) along the 
$\Sigma$ line, between $M$ and $\Gamma$. Furthermore, it did not yield
the second band (II) 
which is almost degenerate with the third band (III) along
$\Sigma$. It is easily shown that the microscopic field 
corresponding to the missing band along $\Delta$ is antisymmetric
with respect to the reflection $G_y\leftrightarrow -G_y$ about
the $\Delta$ line. Similarly,
the missing bands along the region $\Sigma$ are antisymmetric with
respect to the reflection $G_x\leftrightarrow G_y$ about the $\Sigma$
line. Thus, the reciprocal vector $\mathbf 0$ does not contribute to
the electric field for those bands \cite{Robertson}, i.e.,
{\em the missing modes have no 
macroscopic field components} and thus, apparently they may not be
obtained from the roots of the macroscopic dispersion relation
(\ref{TEP}).  

Nevertheless,  within our formalism, the wavevector $\mathbf k$ is
actually not
restricted to lie within the first BZ. Thus, we may calculate the 
macroscopic dielectric function for wavevectors beyond the first BZ,
and obtain the modes in the extended scheme. In
Fig. \ref{f2} we show part of the {\em macroscopic} dielectric
function $\epsilon^M_{T}(\omega,\mathbf k'_\Delta)$ with a
wavevector that has been shifted out of the first BZ from $\mathbf
k_\Delta$ along the $y$ direction by the reciprocal vector
$(2\pi/a)\mathbf y$, i.e., 
$\mathbf k'_\Delta = \mathbf k_\Delta+(2\pi/a)\mathbf y$. We remark
that $\epsilon^M(\omega,\mathbf k)$  is not a periodic function of
$\mathbf k$, as it corresponds to a specific plane wave, not to a
microscopic Bloch's wave. The
intersection of the curve $(k'_\Delta/q)^2$ with
$\epsilon^M_{T}(\omega,\mathbf k'_\Delta)$ 
yields the corresponding {\em macroscopic} modes. For example, the
diamonds in Fig. \ref{f2} illustrate two of the normal modes that
could be obtained using this shifting procedure. One of these modes is
identical to the mode labelled IV that was  obtained
previously using the unshifted response $\epsilon^M_{T}(\omega,\mathbf
k_\Delta)$. Nevertheless, another mode, labelled III appears just
below $qa/2\pi=0.4$ and it corresponds to one of the previously
missing modes. We may shift it back into the 
first BZ in order to display it in the reduced zone
scheme. Proceeding in this fashion, we have obtained all 
of the previously missing bands, shown by diamonds in
Fig. \ref{f3}. Thus, we have shown that {\em we can calculate the full
  photonic band structure} for the TE modes from the macroscopic
non-local dielectric function of the composite system. 

\begin{figure}
  \includegraphics{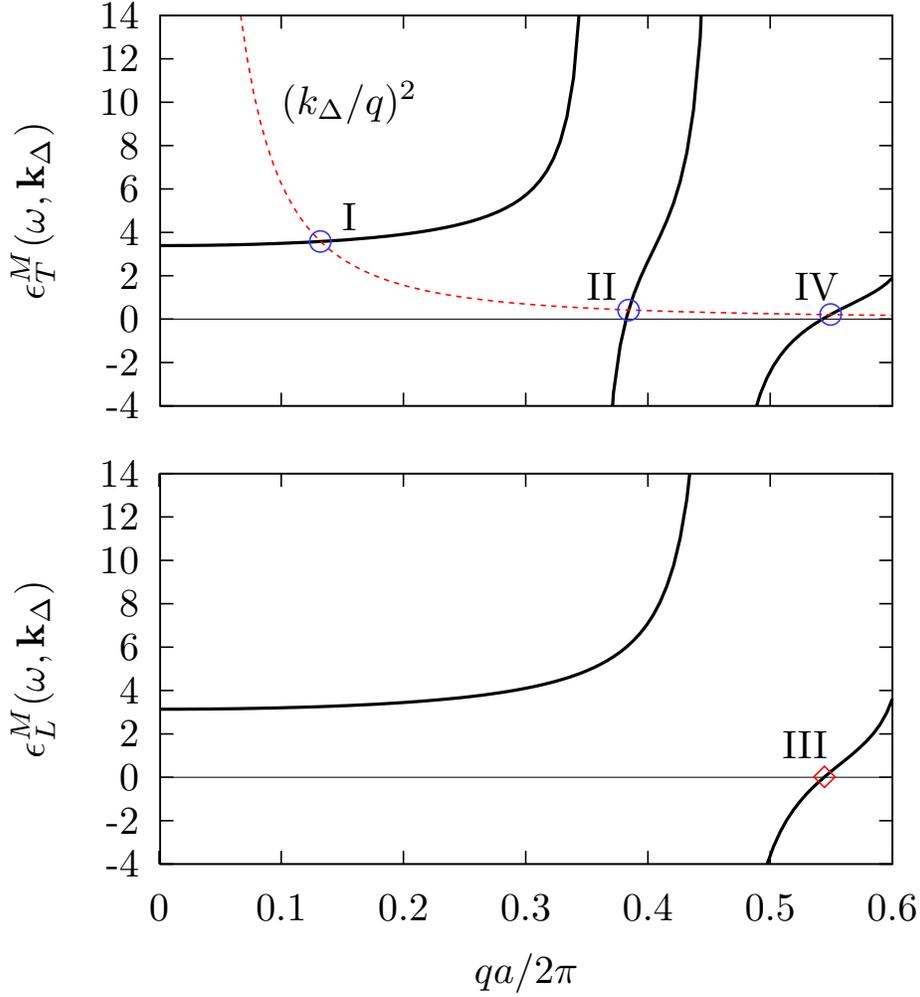}
  \caption{(Color online) In-plane macroscopic dielectric functions
    $\epsilon^M_T(\omega,\mathbf k_\Delta)$ (upper panel) and
    $\epsilon^M_L(\omega,\mathbf k_\Delta)$ (lower panel) of the same 
    system as in Fig. \ref{fig:Eps_M_L1L2} for the same wavevector
    $\mathbf k_\Delta$ as in Fig. \ref{f2}. The line
    $(k_\Delta/q)^2$ (dashes) intersects (circles)
    $\epsilon^M_T(\omega,\mathbf k_\Delta)$ at the
    frequencies $\omega_\alpha(\mathbf k_\Delta)$ of the TM transverse normal
    modes, with $\alpha=$I, II, IV. The zeroes (diamonds)
    of $\epsilon^M_L(\omega,\mathbf k_\Delta)$ yield the longitudinal
    mode III (see text).}
  \label{f4}
\end{figure}

A similar procedure to that discussed above can be employed also for
the TM modes of the system, with the electric field within the $xy$
plane. In Fig. \ref{f4}  
we show the in-plane macroscopic dielectric functions
$\epsilon^M_{xx}(\omega,\mathbf k_\Delta)$ and
$\epsilon^M_{yy}(\omega,\mathbf k_\Delta)$ of the system for the same
wavevector $\mathbf k_\Delta$ as in Fig. \ref{f2}. This wavevector and
the system are symmetric under $y\leftrightarrow -y$ reflections.
Therefore, $\epsilon^M_{xy}=0$ and we may identify the transverse and
longitudinal responses  as $\epsilon^M_T(\omega,\mathbf
k_\Delta)=\epsilon^M_{yy}(\omega,\mathbf k_\Delta)$ and
$\epsilon^M_L(\omega,\mathbf
k_\Delta)=\epsilon^M_{xx}(\omega,\mathbf k_\Delta)$ respectively. The normal modes
are then obtained from the transverse dispersion relation (\ref{TEP})
and the longitudinal dispersion relation \cite{Landau}
\begin{equation}\label{LEP}
  \epsilon^M_L(\omega,\mathbf k)=0.
\end{equation}
These are identified with circles and diamonds in the figure.

\begin{figure}
  \includegraphics{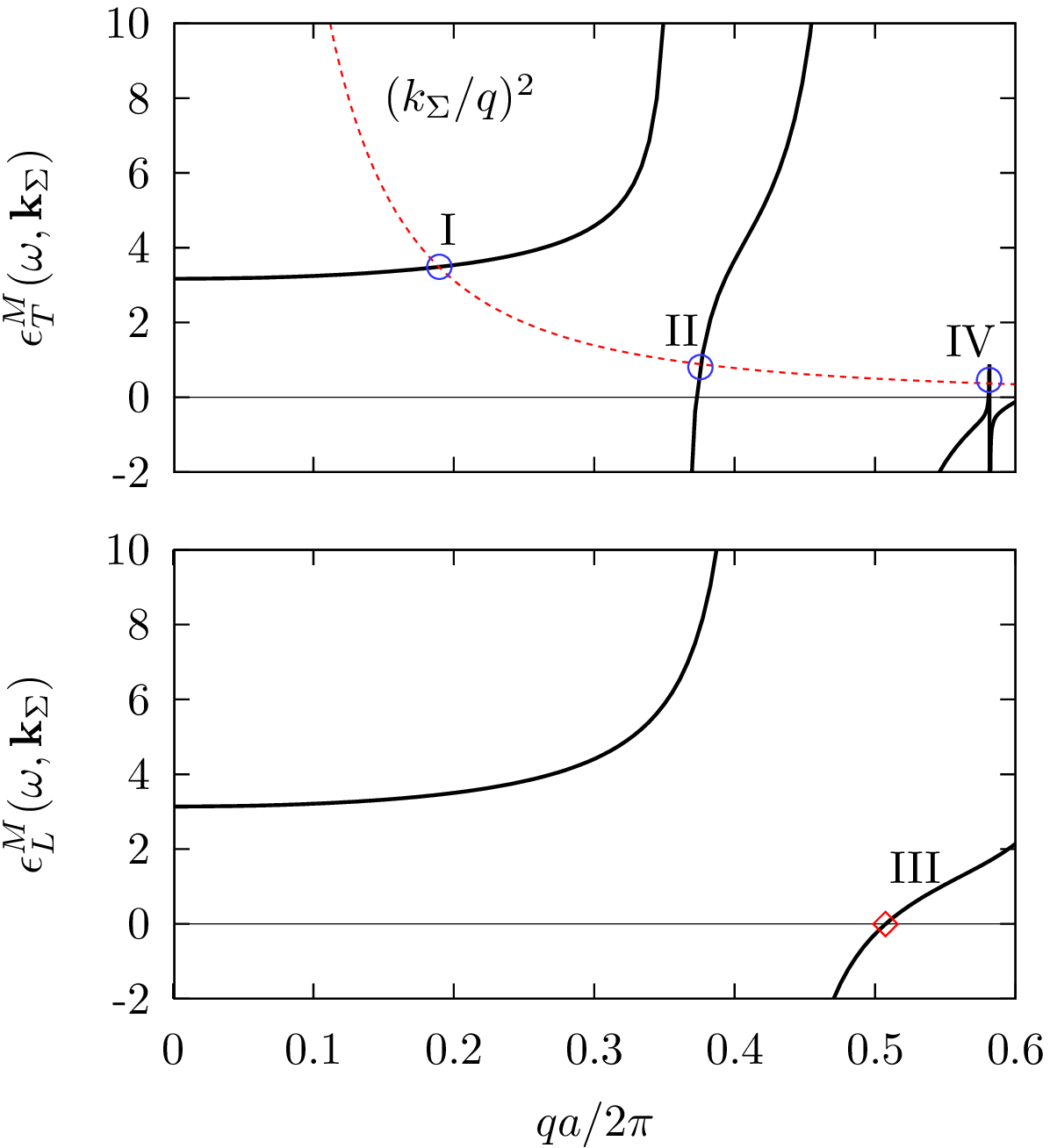}
  \caption{(Color online) In-plane macroscopic dielectric functions
    $\epsilon^M_T(\omega,\mathbf k_\Sigma)$ (upper panel) and
    $\epsilon^M_L(\omega,\mathbf k_\Sigma)$ (lower panel) of the same 
    system as in Fig. \ref{fig:Eps_M_L1L2} for a given wavevector
    $\mathbf k_\Sigma$ along the $\Sigma$ line. The line
    $(k_\Sigma/q)^2$ (dashes) intersects (circles)
    $\epsilon^M_T(\omega,\mathbf k_\Sigma)$ at the
    frequencies $\omega_\alpha(\mathbf k_\Sigma)$ of the TM transverse normal
    modes, with $\alpha$=I, II, IV. The zeroes (diamonds)
    of $\epsilon^M_L(\omega,\mathbf k_\Sigma)$ yield the longitudinal
    mode III (see text).}
  \label{f5}
\end{figure}

In Fig. \ref{f5} we do a similar calculation but along the $\Sigma$
line, for the wavevector $\mathbf k_\Sigma=(\pi/2a,\pi/2a)$. Along
this line, the wavevector and the system are symmetric under the
reflection $x\leftrightarrow y$, and thus $\epsilon^M_{xx}(\omega,\mathbf
k_\Sigma)=\epsilon^M_{yy}(\omega,\mathbf k_\Sigma)$. We may now
identify the transverse and 
longitudinal responses as  $\epsilon^M_T(\omega,\mathbf
k_\Sigma)=\epsilon^M_{xx}(\omega,\mathbf
k_\Sigma)-\epsilon^M_{xy}(\omega,\mathbf  k_\Sigma)$ and
$\epsilon^M_L(\omega,\mathbf 
k_\Sigma)=\epsilon^M_{xx}(\omega,\mathbf
k_\Sigma)+\epsilon^M_{xy}(\omega,\mathbf  k_\Sigma)$  respectively. The upper panel
of Fig. \ref{f5} shows $\epsilon^M_T(\omega,\mathbf k_\Sigma)$,
$(k_\Sigma/q)^2$ and their intersections (circles) which yield the
transverse modes, while the lower panel displays
$\epsilon^M_L(\omega,\mathbf k_\Sigma)$ and its zero (diamond) which
correspond to a longitudinal mode. 

\begin{figure}
  \includegraphics{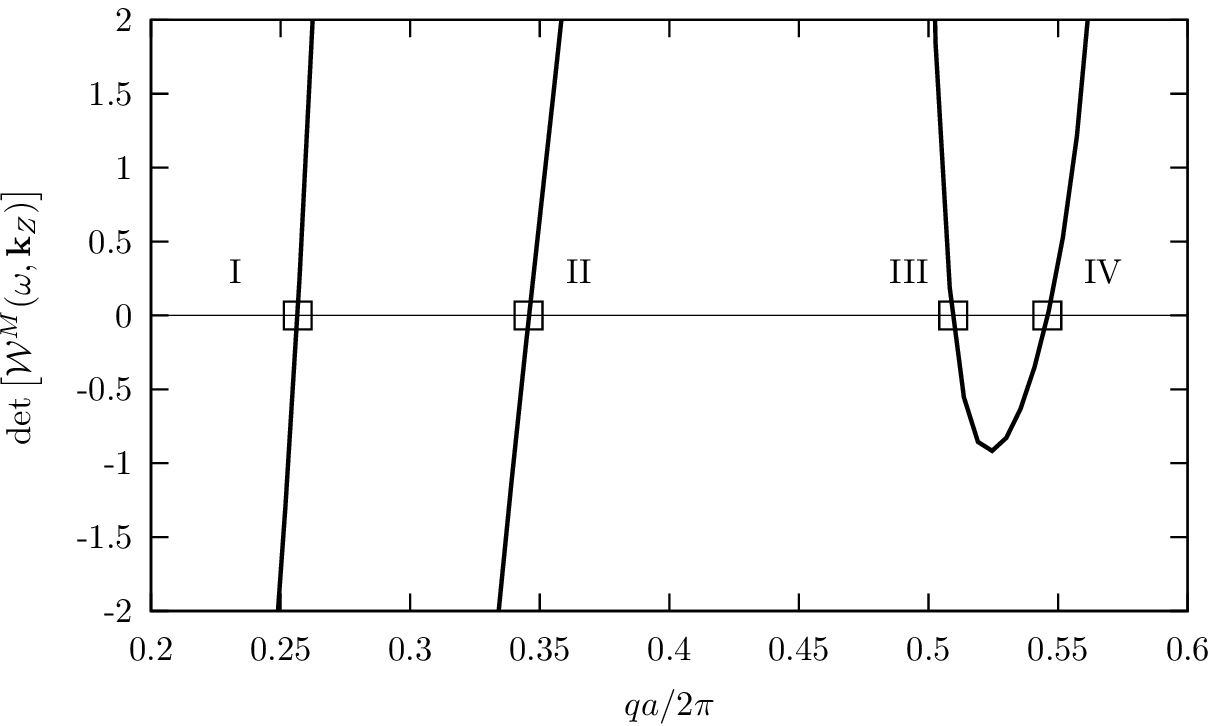}
  \caption{Determinant of the macroscopic wave operator matrix
    $\mathcal W^M(\omega, \mathbf k_Z)$ for a given wavevector
    $\mathbf k_Z$ along the $Z$ line. The normal modes of the system
    $\omega_\alpha(\mathbf k_Z)$, $\alpha=$I, II, III, IV,  correspond to its
    zeroes (squares). } 
  \label{f6}
\end{figure}

Along the $Z$ line, from $X$ to $M$, there is no crystal symmetry
operation that leaves 
invariant the wavevector, and thus transverse and longitudinal fields
mix among themselves, i.e., there are no longitudinal nor transverse
modes. Nevertheless, we may obtain the frequencies of the mixed
modes through the singularities of the wave operator matrix
\begin{equation}\label{detw}
\det\left[\mathcal
W^M(\omega, \mathbf k)\right]=0
,
\end{equation}
 as illustrated in Fig. \ref{f6}
for a wavevector $\mathbf k_Z=(\pi/a, \pi/2a)$ along the $Z$ line. 

\begin{figure}
  \includegraphics{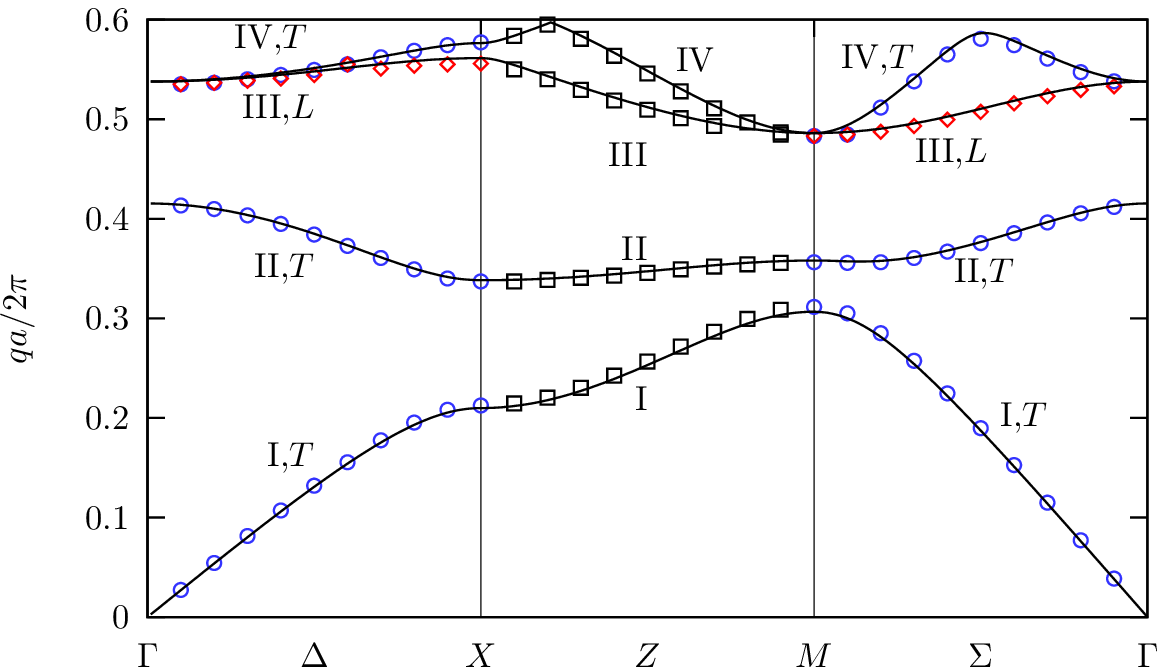}
  \caption{(Color online) Photonic bands of the TM modes of the same
    system as in 
    Fig. \ref{f1}. The modes between $\Gamma$ and $X$ and
    between $M$ and $\Gamma$ may be obtained by
    solving Eq. (\ref{TEP}) for the transverse ($T$) modes (circles) and 
    Eq. (\ref{LEP}) for the longitudinal ($L$) modes (diamonds), or
    Eq. (\ref{detw}) for both of them. The
    modes from $X\to M$ are obtained from Eq. (\ref{detw}) and have a
    mixed polarization (squares). For
    comparison, we also show the photonic bands obtained from the
    solution of     the corresponding eigenvalue problem, 
    Eq. (\ref{TMMP}), as described in the text (solid lines).
}
\label{f7}
\end{figure}

From calculations such as those illustrated by Figs. \ref{f4},
\ref{f5} and \ref{f6}, or more generally, from the zeroes of $\det
\left[\mathcal W^M(\omega,\mathbf k)\right]$ for arbitrary wavevectors $\mathbf k$,
we can obtain the full TM photonic band structure, as shown by
Fig. \ref{f7}.
In order to test our theory, we also plot the TM modes
obtained by solving the eigenvalue equation \cite{Joannopoulos}
\begin{equation}\label{TMMP}
\nabla\cdot\frac{1}{\epsilon(\mathbf r)}
\nabla
B_z(\mathbf r)= -
  \frac{\omega^2}{c^2}
B_z(\mathbf r)
\end{equation}
using similar methods \cite{Saad, unpublished} as for the TE case and comparing them
successfully to previous results \cite{Busch} for the same system. The
agreement between 
these calculations shows that it is also feasible to use the macroscopic
dielectric response of the system to obtain its photonic band
structure in the TM case.

\begin{figure}
    \includegraphics{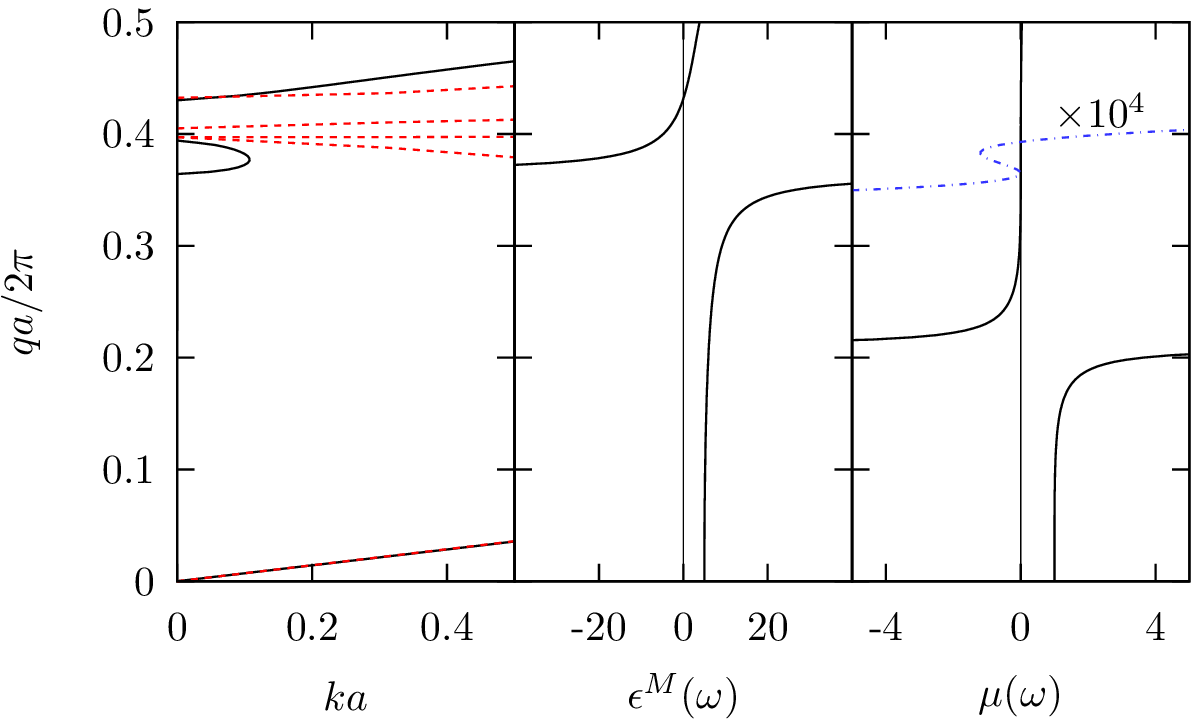}
  \caption{\label{f8}Local dispersion relation for the same system as
    in Fig. \ref{f1} for the TE case (solid lines, left panel),
    obtained from Eq. (\ref{local}). Local
    dielectric function $\epsilon^M(\omega)$ (middle
    panel). Local magnetic permeability $\mu(\omega)$ (solid line, right
    panel) and $10^4\mu(\omega)$ (dash-dotted line, right panel). For
    comparison purposes, we also show the exact dispersion 
  relation (dashed line, left panel).}
  
\end{figure}

An approximate band structure could be obtained for transverse waves
in the region of small  wavevectors
$\mathbf k\to \mathbf 0$ by expanding the LHS of Eq. (\ref{TEP}) up to
second order in $k$ and solving for $k^2$. The result is a local
dispersion relation
\begin{equation}\label{local}
  k^2=q^2 \epsilon^M(\omega)\mu(\omega),
\end{equation}
where the non-locality of $\epsilon^M_T(\omega,\mathbf k)$ is
partially accounted for through an effective {\em magnetic
  permeability}  \cite{Costa}
\begin{equation}\label{mu}
  \mu(\omega)= \left
  . \frac{1}{1-\frac{q^2}{2}\frac{\partial^2}{\partial 
      k^2}\epsilon^M_T(\omega, k \mathbf x)} \right |_{k=0}.
\end{equation}
In Fig. \ref{f8} we show the resulting approximate bands for the TE
case. 
We can see that the acoustic band within the local approximation is
indistinguishable from the exact result for small wavevectors. This local
acoustic band
extends up to the frequency $qa/2\pi\approx 0.2$ for
which $\mu$ has a pole. The local approximation also reproduces the
upper band of the exact calculation (labelled V in Fig. \ref{f3}), but
only in the immediate vicinity of $\Gamma$, close to the zero
$qa/2\pi\approx0.43$ of
$\epsilon^M(\omega)$. For larger wavevectors it acquires a larger
positive dispersion. As discussed above, the band labelled III in
Fig. \ref{f3} doesn't couple to macroscopic fields for $\mathbf k$
along the $\Delta$ line, while the band labelled IV doesn't couple
along the $\Sigma$ line. Thus, neither band couples to macroscopic fields
at $\Gamma$ and as a consequence, these bands are not reproduced at all by
the local approximation (\ref{local}). 

In the region $qa/2\pi\approx0.36-0.40$ the
local approximation predicts a curious band that displays
backbending. Its lower part begins at the pole of
$\epsilon^M(\omega)$, which corresponds to a double zero
of $\mu(\omega)$ \cite{unpublished}. This
part of the local dispersion relation is spurious, as the second order
Taylor expansion of the dielectric function that yields
Eqs. (\ref{mu}) and (\ref{local}) starting from Eq. (\ref{TEP}) would
be meaningless at a pole. Consequently, this part of the local band doesn't
correspond to any band within the exact calculation. On the other
hand, the upper part of the backbending band starts at a simple
zero of $\mu$ which arises from a different kind of singular behavior,
consisting of a pole in $\epsilon^M_T(\omega,\mathbf k)$ at $qa/2\pi\approx
0.40$ which is 
suppressed at $k=0$ as its weight is approximately proportional to
$k^2$. This part of the local dispersion relation agrees
with band II of Fig. \ref{f3} at $\Gamma$  {\em and displays a negative
  dispersion} as does band II. Furthermore, for this band, both
$\epsilon^M$ and $\mu$ are negative. Thus, our photonic crystal, made
up of holes within a dispersionless dielectric, behaves for some
frequencies as a left-handed metamaterial
\cite{Peng,Vynck}. Nevertheless, caution should be exercised when
using Eqs. (\ref{mu}) and (\ref{local}) close to singularities of
$\epsilon^M_T(\omega,\mathbf k)$.

\section{Conclusions}
\label{conclusion}
We have shown that the macroscopic inverse wave operator of composite
systems with arbitrary spatial fluctuations is simply given by the
average projection of the corresponding microscopic operator. These
operators may be related to the macroscopic and macroscopic dielectric
functions, 
thus yielding a general 
procedure to incorporate the effects of spatial 
fluctuations in the calculation of all the components of the
macroscopic dielectric tensor of the system. We have extended
Haydock's recursive scheme in order to calculate very efficiently the
response of periodic two-component systems in terms of a continued
fraction whose coefficients may be obtained without recourse to
operations with the large matrices that characterize the microscopic
response and the field equations. We illustrated and tested our
results through the calculation of the wavevector and frequency
dependent macroscopic dielectric tensor of a 2D dielectric system
with non-dispersive non-dissipative components. We identified the
transverse and longitudinal components of the response for wavevectors
along symmetry lines and we obtained all the components of the
tensorial response for the case of reduced symmetry. The macroscopic
response has a series of poles, related to resonant multiple coherent
reflections and by accounting for its spatial dispersion
we obtained the full photonic band structure of
the system from the dispersion relations corresponding to the
homogenized material. Besides yielding the correct bands from a
macroscopic approach, our scheme allowed us to classify the
polarization of each mode as either transverse, longitudinal or
mixed. Even though the macroscopic approach might fail to yield some
modes, namely, those which have no coupling to the macroscopic field due to
symmetry, they may be recovered by extending the notion of macroscopic
state, allowing its wavevector to lie beyond the first Brillouin zone.
The non-locality of the dielectric response may be partially accounted for
through a magnetic permeability which can then be employed in the
calculation of the optical properties of the system. We compared the
band structure obtained through this local approximation to the exact
results and we obtained partial agreement close to $\Gamma$ for those
bands that do couple to long-wavelentgh fields. In particular, we
showed that this approach can yield a negative dispersion in frequency
regions where both the permitivity and permeability are
negative. Nevertheless, we discussed some shortcommings of the local
approach.  Although here we presented results for dispersionless
transparent dielectrics, our
calculation does not require that the 
materials that make up the system be non-dispersive nor
non-dissipative, so that calculations for real dielectrics and metals
may be performed with the same low computational costs. 

\ack JSPH is grateful for a scholarship awarded by  CONACYT. This work
was partially supported by ANPCyT grant 
PICT-PRH-135-2008 (GPO), CONACyT grant 153930 (BSM) and by DGAPA-UNAM
grant IN108413 (WLM). 

\section*{References} 
\bibliographystyle{unsrt}
\bibliography{Haydockbib}

\end{document}